\documentclass[
 aip,
 jmp,
 amsmath,amssymb,
 reprint,
]{revtex4-1}
\usepackage{graphicx}
\usepackage{dcolumn}
\usepackage{bm}
\usepackage{color}

\begin{document}

\title{Shear-wave generation from cavitation in soft solids}

\author{J. Rapet}
 \email{JULIEN001@e.ntu.edu.sg}
 \affiliation{ Division of Physics and Applied Physics, School of Physical and Mathematical Sciences, Nanyang Technological University, 21 Nanyang Link, 637371, Singapore}
 \affiliation{
 	Institute of Physics, Otto-von-Guericke Universit\"{a}t Magdeburg, Universit\"{a}tsplatz 2, 39016 Magdeburg, Germany
 }
\author{Y. Tagawa}
 \affiliation{Department of Mechanical Systems Engineering, Tokyo University of Agriculture and Technology, Naka-cho 2-24-16, Koganei, Tokyo 184-8588, Japan
 }
\author{C.D. Ohl}
 \affiliation{
  Institute of Physics, Otto-von-Guericke Universit\"{a}t Magdeburg, Universit\"{a}tsplatz 2, 39016 Magdeburg, Germany
 }
  \affiliation{ Division of Physics and Applied Physics, School of Physical and Mathematical Sciences, Nanyang Technological University, 21 Nanyang Link, 637371, Singapore}

\date{\today}

\begin{abstract}
The formation and dynamics of cavities in liquids leads to focusing of kinetic energy and emission of longitudinal stress waves during the cavity collapse. 
Here we report that cavitation in elastic solids may additionally emit shear waves that could affect soft tissues in human bodies/brains.
During collapse of the cavity close to an air-solid boundary, the cavity moves away from the boundary and forms a directed jet flow, which confines shear stresses in a volume between the bubble and the free boundary. Elastographic and high-speed imaging resolve this process and reveal the origin of a shear wave in this region. Additionally, the gelatin surface deforms and a conical crack evolves. We speculate that tissue fracture observed in medical therapy may be linked to the non-spherical cavitation bubble collapse.  
\end{abstract}

\maketitle
In many medical applications energy is deposited within the tissue resulting in the formation of a cavity which expands and collapses. 
This phenomena is termed cavitation. The understanding of its dynamics helps to improve precision and mitigate side effects~\cite{Vogel2005Nanosurgery,brennen2015cavitation}. 
Examples of cavitation based therapy are histotripsy where pulsed finite amplitude ultrasound waves are focused into tissues and cornea surgery where laser pulses locally vaporize corneal tissue and produce clean intrastromal cuts \cite{juhasz1999corneal, lubatschowski2000application}. The former leads to regions of intense cavitation where tissue is rapidly and locally turned into a paste\cite{roberts2006pulsed, kim2011non}. The importance of bubble-tissue interaction stimulated research on individual bubbles oscillating in a tissue mimicking elastic solid. There the emission of stress and tension waves during bubble generation and collapse were documented~\cite{brujan2006stress}.

Tissue mimicking materials such as gelatin, polyacrylamide gels (PAA) or other hydrogels provide an elastic restoring force. Thus, they can transport not only longitudinal waves (with a wave velocity of $\approx 1500\,$m/s) but also transversal waves propagating at a considerably smaller velocity of $\approx 1-50$ m/s. As the shear wave velocity depends on the mechanical properties of the tissue, the tissue type may be characterized in diagnostic ultrasound from the shear front propagation~\cite{gennisson2013ultrasound}. 
Beside the established methods based on acoustic focusing\cite{song2012comb}, shear wave generation is an active research field with a number of new actuation mechanisms being reported. For example using an electric current in combination with a magnetic field displaces the elastic solid by the Lorentz force\cite{grasland2014imaging} or by locally displacing the hydrogel with bubbles from electrolysis\cite{montalescot2016electrolysis}. Recently, the generation of shear waves from local heating was demonstrated: when a laser beam heats up the surface of a soft elastic absorber\cite{grasland2016generation} two regimes are observed. In the thermoelastic regime the waves arise from the thermal expansion and in the ablative regime from impulse transfer due to vaporization of the surface.

Shear waves of larger amplitude, often induced by impacts\cite{cooper1989biophysics, taylor2014investigation}, cause the negative effects on soft tissues, e.g. in blast-induced traumatic brain injuries\cite{taylor2014investigation,taber2006blast}.
In the present work, we report on mechanism of shear wave generation from the  non-spherical collapse of a single cavitation bubble in a soft solid. To induce the non-spherical collapse, the bubble is created in the solid near an air-solid boundary. 

\begin{figure}
	\includegraphics[width = 0.5\textwidth]{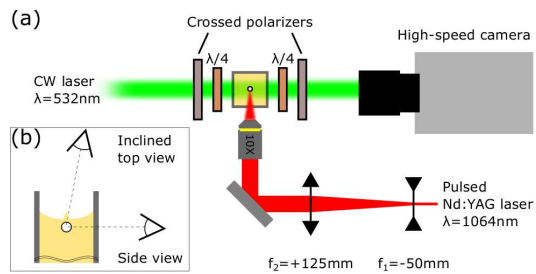}
	\caption{\label{fig:Setup} Experimental setups for bubble generation and observation. (a) Top view of the setup with the circular polariscope. The recording path is shown in green and the bubble generation path in red. The polarizers and $\lambda$/4 plates are used for visualizing shear waves. (b) Side view of the setup, without the polariscope, showing the orientation of the cameras used for synchronized recording of the bubble and surface dynamics.
	}
\end{figure}

\begin{figure*}
\includegraphics[width = 1\textwidth]{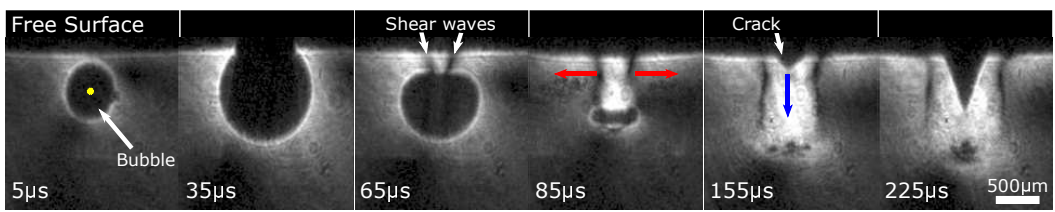}
\caption{\label{fig:Pola-side} Selected images of the dynamic of a bubble nucleated in gelatin near a free surface (on top) observed using a circular polariscope. The yellow dot at 5 $\mu$s indicates nucleation site, red arrows at 85 $\mu$s show the shear waves propagation directions, and a blue arrow at 155 $\mu$s reflects the crack dynamics.}
\end{figure*}

High-speed photography is used to record the bubble dynamics and the shear wave propagation. The latter is compared with a simple model to support the hypothesis of the mechanism of shear wave generation.

While pressure waves typically result in a change of index of refraction and can be visualized with shadowgraphy or Schlieren imaging, shear waves are more difficult to picture. Here, we utilize birefringence in the solid. Gelatin as our tissue phantom becomes doubly refractive under stress and is then suitable for photoelastic photography \cite{tomlinson2015photoelastic}. 
Those two refracted rays possess directions of polarization coinciding with the local principal stress directions. A circular polariscope is particularly suitable for measuring the stress distribution in the medium. It is obtained from two linear polarizers and two $\lambda/4$ waveplates as sketched in Figure ~\ref{fig:Setup}(a). 
The intensity of the light exiting the circular polariscope is a function of the retardation $\delta$ (Eq.~\ref{eq:retardation}) and can be expressed using trigonometry or Jones calculus \cite{ramesh2000digital} for 2-dimensional stress distributions
\begin{equation}\label{eq:circ-pola}
I = k^2\,\sin^2\bigg(\frac{\delta}{2}\bigg)\quad .
\end{equation} 
Monochromatic illumination assures that the relative retardation $\delta$ is only a function of the difference in amplitude of the two principal stresses and can be expressed as
\begin{equation}\label{eq:retardation}
\delta = \frac{2\pi h}{\lambda}C(\sigma_1-\sigma_2)\quad ,
\end{equation}
where $h$ is the thickness of the sample, $\lambda$ the wavelength of the incoming light, $C$ the stress-optic coefficient, $\sigma_1$ and $\sigma_2$ the principal stresses.
Thus the light is extinguished only where the sample is unstressed or where the principal stress difference $(\sigma_1 - \sigma_2)$ is causing a phase difference of $\delta = 2m\pi$ $ (m = 0,1,2,...)$. 
The second condition marks the presence of an isochromatic, i.e an area of constant principal stress difference.

The experimental setup shown in Figure ~\ref{fig:Setup}(a) allows to record the cavitation bubble dynamics and the shear wave emission through photoelastic imaging with a high speed camera.
Single laser induced cavitation bubbles are created by focusing a laser pulse (Litron Lasers, Nano series, Q-switched Nd:YAG, 6 ns, wavelength $1064\,$nm ) into the gelatin with a microscope objective (Olympus $10\times$ Plan Achromat, N.A. $= 0.25$).
At the focal point of the lens the bubble is created through an optical breakdown. The dynamics of the bubble (expansion, collapse and rebound) and the shear waves are recorded with a high speed camera (Shimadzu HPV-X2) equipped with a macro lens (Canon MP-E 65mm f/2.8 1-5$\times$ Macro). Single wavelength illumination is provided by a continuous green laser (Shaan'’xi Richeng Ltd, DPSS Green Dot Laser Module, wavelength $532\,$nm).

The soft solids samples are prepared from powdered gelatin (Gelatin 250 bloom, Yasin Gelatin CO.,LTD). It is mixed with deionized water at a mass ratio of 4\% (gelatin to water) and  dissolves in a flask on a hot plate with a magnetic stirrer. The hot mixture is poured into an optical glass cuvette (Aireka Scientific Co. Ltd, $3.5\,$ml, QO10204-4) to minimize birefringence distortion from the container and insure a flat gelatin-glass interface for clear observation and accurate laser focusing. 
The samples cool down to room temperature and are stored in a fridge. Before use, we assure that the samples have reached room temperature again.  

\begin{figure*}
	\includegraphics[width = 1\textwidth]{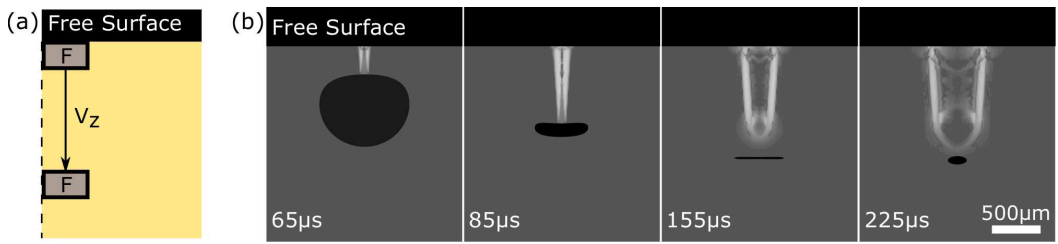}
	\caption{\label{fig:Simu} Finite element simulation of the wave generation. (a) Simulation domain. The region $F$ indicates a rectangular source of force moving downward from the free surface along the axis of symmetry with a velocity of $V_Z$. (b) Selected frames of the numerical results after light intensity calculation.}
\end{figure*}

Figure~\ref{fig:Pola-side} presents a typical result  from the experimental setup Figure ~\ref{fig:Setup}(a). It consists of selected frames from a high-speed recording of a bubble expanding and collapsing in gelatin near a free (air) boundary. The time $t = 0$ denotes the time of bubble nucleation. Then the bubble expands and compresses the gelatin nearby resulting in a bright area around the bubble ($t = 5\,\mu$s). Upon reaching the maximum size, the bubble pushes and deforms the free surface ($t=35\,\mu$s) before it shrinks again. Thereby its upper wall bulges in and forms a funnel while the position of the interface at the bottom of the bubble remains approximately static ($t= 65\,\mu$s). Between the bubble and the free surface, two vertical dark lines develop on top of the brighter stressed area. These two lines mark the presence of isochromatics. 
The dark lines extend vertically following the bubble collapse with a velocity of $V_v = 25\,$m/s$\,\pm\,1$ m/s, which roughly also is the velocity of the upper bubble interface. Their horizontal position changes, too. They move outward with a velocity of $V_s =2\,$m/s$\,\pm\,0.2\,$m/s.

After reaching the minimum size the bubble rebounds ($t = 85\,\mu$s) and its volume oscillations cease shortly after the second collapse at $t = 119\,\mu$s (not shown), while it remains moving downwards.

At $t = 155\,\mu$s, a conical-shaped crack appears starting from the air-gelatin boundary and follows the path of the downwards translating bubble ($t = 225\,\mu$s). We speculate that the shrinking and moving bubble is the source for shear stress generation. As the speed of this motion is considerably higher than the shear wave velocity, shear stress is confined in the region bounded by the air interface and the bubble. Later only this confined stress propagates from this region as a shear wave into the surrounding medium.

This hypothesis is tested with a numerical simulation of the stress generation and propagation using finite element solver for an elastic solid (Solid Mechanics Module, COMSOL Multiphysics). For the sake of simplicity, the stress field resulting from a body force moving downward with a constant velocity is modeled. Figure~\ref{fig:Simu}(a) depicts the 2D axisymmetric simulation domain, properties of the gelatin are taken from literature ($\rho = 1000$ kg/m$^3$, $\epsilon = 0.45$, $E = 10$ kPa, $C_p = 1500$ m/s, $C_s = 1.8$ m/s){\cite{czerner2015determination,gennisson2013ultrasound}}. 
The boundary conditions are stress free on the top, an axis of symmetry, and to the right and bottom of the domain impedance matched boundaries. 
At $t = 0$, a rectangular source of force (body load, $F = 1\,$N) moves downward from the free surface along the axis of symmetry with a velocity of $V_Z = 25$ m/s for a duration $\Delta t = 32\,\mu$s. To compare the simulations with the photoelastic images we split the axisymmetric model into slices perpendicular to the line of sight. 
In each slice of finite thicknesses, the polariscope equation is solved assuming a constant stress along the depth. 

Figure ~\ref{fig:Simu}(b) presents selected frames from the numerical results. 
Black areas correspond to the gas/vapour domains (bubble and free surface drawn from the experimental results and added on the numerical results), dark gray (low light intensity) shows the unstressed gelatin and areas of stress are depicted with gray scales, with higher brightness indicating areas of higher stress. 
Overall the simulation results show good agreement considering the coarse simplifications. In particular we observe a bright region which extend vertically following the rectangular source of force with a velocity of $V_z$ and slowly expands radially. The main difference between simulation and experiment is that the isochromatics are not reproduced. We explain this that the pre-existing stress fields from the earlier bubble dynamics are not considered in the simulations, i.e. there the body force moves within an unstressed solid. The simulation supports the explanation that shear waves can be induced with a source of stress moving with the bubble upper interface. 

\begin{figure*}
	\includegraphics[width=1\textwidth]{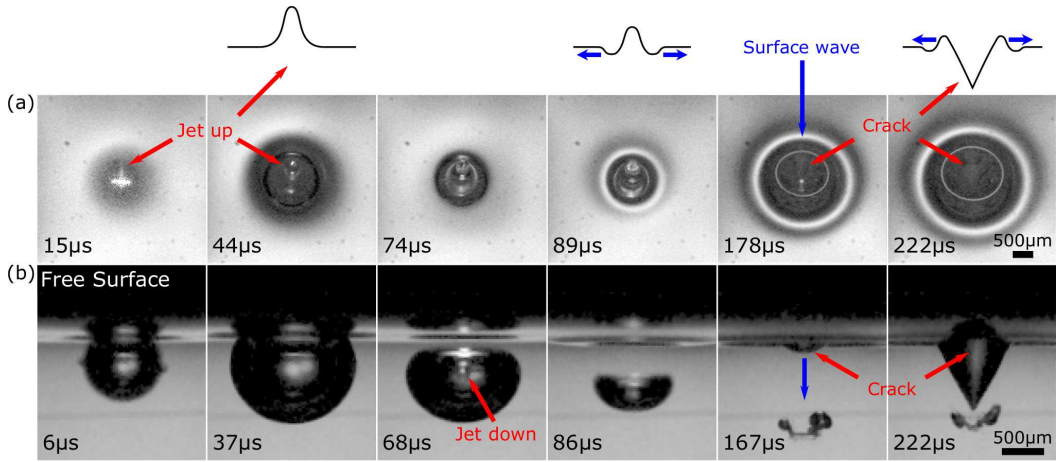}
	\caption{\label{fig:2-cam}  Selected images of the dynamic of a bubble nucleated in gelatin near a free surface taken by the setup shown in Figure 1(b). (a) Top view. (b) Side view. The blue arrow reflects the crack dynamics.
    Frames were taken from a single experiment and were selected to show the event at similar time step. The variation is due to the different recording speeds ((a), $67.5$ kfps - (b), $162$ kfps). Sketches above show the profile of the air-gelatin interface for some of the presented frames.}
\end{figure*}

The motion of the bubble collapsing away from the free surface (Figure \ref{fig:Pola-side}) is also reported for liquids, i.e. in water where a jet pierces through the bubble \cite{robinson2001interaction}. 
It suggests that the emergence of a jet penetrating through the bubble in gelatin could cause the generation of the shear waves.
To obtain a better photographic evidence we repeated the experiment in absence of the polariscope optics, see Figure ~\ref{fig:Setup}(b) for the experimental setup. Now two high-speed cameras (Photron, Fastcam, Mini AX200) equipped with camera lenses (Canon MP-E 65 mm f/2.8 1-5$\times$ Macro) record simultaneously the bubble and the surface from the side and top.

Figure~\ref{fig:2-cam} presents selected frames from a bubble with very similar size and position as shown in Figure \ref{fig:Pola-side}. Again the bubble expands and pushes the free surface resulting in an upward displacement visible on the surface with a gelatin jet pointing upward ((a):$t = 15\,\mu$s) and expanding laterally as the bubble grows ((a):$t = 44\,\mu$s, (b):$t = 37\,\mu$s).  
The bubble then collapses and moves away from the free surface. During collapse, a jet is penetrating through  the center of the bubble with a velocity of $V_s = 19\,$m/s$\,\pm\,1\,$m/s ((b):$t = 68\,\mu$s).
Then the jet impacts on the lower bubble interface while the bubble continues to translate downward ((b):$t = 86\,\mu$s and $t= 168\,\mu$s).
These observations indicate that the jet formed by the bubble moving away from the air-gelatin boundary is causing the shear wave.
At a later time the restoring force of the gelatin pulls back the residues of the bubble towards its position at creation, see (b):$t = 222\,\mu$s.

The crack originating from the air-gelatin boundary indicates that the upwards pointing tip of the jet on the gelatin-air interface ((a):$t = 15 \,\mu$s) retracts first towards the original level and then penetrates into the gelatin, visible as a conical crack in (a):$t = 178\,\mu$s. The opening angle of the crack remains while the crack tip propagates into the gelatin trailing the cavitation bubble (Figure \ref{fig:Pola-side}(b):$t = 222\,\mu$s). The oscillations of the free surface result in surface waves (Figure ~\ref{fig:2-cam} sketch and (a)) traveling radially with a velocity of $V_s =  4.5\,$m/s$\,\pm\,0.5\,$m/s.

In conclusion, we have demonstrated that a single cavitation bubble in soft media can generate shear waves. The jetting of the shrinking and collapsing bubble is causing shear stresses to be confined between the air-gelatin interface and the bubble. Shear waves are then emitted from this region. We also observe the formation of a crack following the downward motion of the bubble.
We speculate that this could be a potential source of tissue damage when cavitation bubbles collapse non-spherically. 
We speculate that confinement of shear stress near to free interfaces may lead to cracks and tissue damage. 

\textit{Acknowledgements} Y.T. acknowledges financial support from JSPS KAKENHI Grants No. R2801 and 17H01246. This project has received funding from the European Union Horizon 2020 Research and Innovation programme (No 813766).

\bibliography{Bib_shear}

\end{document}